\documentclass[11pt]{article}

\begin{document}

\title{CONSISTENT INTERACTIONS IN A THREE-DIMENSIONAL THEORY WITH TENSOR GAUGE
FIELDS OF DEGREES TWO AND THREE}
\author{SOLANGE--ODILE SALIU\thanks{%
E-mail: osaliu@central.ucv.ro} \\
Faculty of Physics, University of Craiova, 13 Al. I. Cuza Street\\
Craiova, RO-1100, Romania}
\maketitle

\begin{abstract}
All consistent interactions in a three-dimensional theory with tensor gauge
fields of degrees two and three are obtained by means of the deformation of
the solution to the master equation combined with cohomological techniques.
The local BRST cohomology of this model allows the deformation of the
Lagrangian action, accompanying gauge symmetries and gauge algebra. The
relationship with the Chern--Simons theory is discussed.

\end{abstract}

\section{Introduction}

The matter of generating consistent interactions in gauge theories~\cite
{alpha1}--\cite{alpha3} has been reanalyzed as a deformation problem of the
master equation~\cite{def} in the framework of the antifield-BRST formalism~%
\cite{1}--\cite{5}. The scope of this paper is to generate all consistent
interactions that can be introduced in a free three-dimensional theory with
tensor gauge fields of degrees two and three by means of the deformation of
the solution to the master equation combined with cohomological techniques.
The importance of the three-dimensional case is motivated by the fact that
the model is irreducible, and thus its deformation displays different
features as compared to the higher dimensional situations, where the theory
is reducible (the reducibility order depends on the space-time dimension).
The model under consideration can also be interpreted in terms of a special
class of theories with mixed-symmetry type tensor gauge fields. Such
theories held the attention lately on many issues, like, for example, the
interpretation of the construction of the Pauli--Fierz theory~\cite{pf}, the
dual formulation of linearized gravity~\cite{dual}--\cite{lingr}, the
impossibility of consistent cross-interactions in the dual formulation of
linearized gravity~\cite{lingr}, or the general scheme for dualizing
higher-spin gauge fields in arbitrary irreducible representations of $GL(D,%
\mathbf{R})$~\cite{gensch}. Meanwhile, the model under study is connected to
a special class of topological BF theories~\cite{topBF}, and also to the
pure Chern--Simons theory.

Our approach is based on solving the main equations that govern the
deformation of the solution to the master equation for the starting free
model. Since the model is Abelian and irreducible, its antibracket-antifield
BRST symmetry reduces to the sum between the Koszul--Tate differential and
the exterior derivative along the gauge orbits. Initially, we compute the
non-integrated density of the first-order deformation of the solution to the
master equation, which lies in the local cohomological space of the BRST
differential at ghost number zero. This step relies on the development of
the BRST co-cycles according to the degree of the Koszul--Tate differential
(antighost number), and necessitates the computation of both the cohomology
of the exterior derivative along the gauge orbits and the local homology of
the Koszul--Tate differential. We show that the first-order deformation
stops at antighost number two and is parametrized in terms of seven
independent real constants. The consistency of the first-order deformation
restricts the number of independent constants to three, and consequently
ends the deformation procedure at order one in the coupling constant. The
resulting deformations are classified according to three complementary
classes. Irrespective of the considered class, the interacting theory
exhibits the following general features: (a) the interaction vertices are
polynomials of order three in the undifferentiated tensor gauge fields; (b)
the gauge transformations of the fields are modified; (c) the deformed gauge
algebra is non-Abelian ($su(2)$), unlike that of the initial theory, which
is Abelian. Finally, we discuss the link with pure three-dimensional
Chern--Simons theory.

\section{Free Model. Free BRST Symmetry}

We begin with the Lagrangian action in three space-time dimensions
\begin{equation}
I_{0}\left[ B_{\;\;\;(\lambda )}^{\alpha \beta },A_{\alpha }^{\;\;(\lambda
)}\right] =\int d^{3}xB_{\;\;\;(\lambda )}^{\alpha \beta }\partial _{\alpha
}A_{\beta }^{\;\;(\lambda )},  \label{bf31}
\end{equation}
that involves two sorts of covariant tensor gauge fields: one of degree
three, which is only antisymmetric in its first two indices, $%
B_{\;\;\;(\lambda )}^{\alpha \beta }=-B_{\;\;\;(\lambda )}^{\beta \alpha }$,
and the other of degree two, $A_{\alpha }^{\;\;(\lambda )}$, with no further
symmetries. We use the Minkowskian metric of `mostly minus' signature, $%
\left( +,-,-\right) $, and work with the convention that the completely
three-dimensional symbol $\varepsilon ^{\alpha \beta \gamma }$ is valued
like $\varepsilon ^{012}=+1$. All indices ($\alpha $, $\beta $, $\lambda $)
are lowered and raised with the Minkowski metric. Parentheses around an
index are meant to mark that there is no symmetry between the class of
indices carrying parentheses and that without. In terms of Young diagrams, $%
B_{\;\;\;(\lambda )}^{\alpha \beta }$ can be regarded like the tensor
product between a two-cell diagram with one column and a one-cell diagram,
while $A_{\alpha }^{\;\;(\lambda )}$ is simply the tensor product between
two one-cell diagrams
\begin{equation}
B_{\alpha \beta (\lambda )}\simeq
\begin{tabular}{|l|}
\hline
$\alpha $ \\ \hline
$\beta $ \\ \hline
\end{tabular}
\bigotimes
\begin{tabular}{|l|}
\hline
$\lambda $ \\ \hline
\end{tabular}
,\;A_{\alpha (\lambda )}\simeq
\begin{tabular}{|l|}
\hline
$\alpha $ \\ \hline
\end{tabular}
\bigotimes
\begin{tabular}{|l|}
\hline
$\lambda $ \\ \hline
\end{tabular}
.  \label{young}
\end{equation}
In other words, the representations of $GL\left( 3,\mathbf{R}\right) $ in
the corresponding spaces of covariant tensors of degree three and
respectively two are reducible. The gauge symmetries of action (\ref{bf31})
are `spanned' by the irreducible generating set
\begin{equation}
\delta _{\epsilon }B_{\;\;\;(\lambda )}^{\alpha \beta }=\varepsilon ^{\alpha
\beta \gamma }\partial _{\gamma }\epsilon _{(\lambda )}^{\prime },\;\delta
_{\epsilon }A_{\alpha }^{\;\;(\lambda )}=\partial _{\alpha }\epsilon
^{(\lambda )},  \label{bf32}
\end{equation}
that closes according to an Abelian algebra, with $\epsilon _{(\lambda
)}^{\prime }$ and $\epsilon ^{(\lambda )}$ two independent real and bosonic
covariant tensors of degree one. Although the parentheses might seem
superfluous at degree one, we will still keep them for subsequent analysis.
The above action describes a gauge theory of Cauchy order two. In the case
where the covariant index of the type $(\lambda )$ is replaced by one of
different kind, say $a=\overline{1,N}$, the corresponding action (\ref{bf31}%
) relates to a particular class of topological BF-type theories~\cite{topBF}
involving a system of $U(1)$-Abelian vector fields $A_{\mu }^{a} $ and a set
of Abelian two-form gauge fields $B_{a}^{\mu \nu }$. If, in addition, the
two-form is dual to the one-form, $B_{a}^{\mu \nu }=\left( 1/2\right)
k_{ab}\varepsilon ^{\mu \nu \lambda }A_{\lambda }^{b}$ (with $k_{ab}$ the
elements of a constant, symmetric and invertible matrix), then action (\ref
{bf31}) reduces to a sum of $N$ Abelian Chern--Simons terms. In order to
display the formulas in a simpler manner, it is convenient to replace $%
B_{\;\;\;(\lambda )}^{\alpha \beta }$ by its dual with respect to the
antisymmetry indices only, $V_{\alpha (\lambda )}=\varepsilon _{\alpha \beta
\gamma }B_{\;\;\;(\lambda )}^{\beta \gamma }$, in terms of which (\ref{bf31}%
) and (\ref{bf32}) become
\begin{equation}
I_{0}\left[ V_{\alpha (\lambda )},A_{\alpha }^{\;\;(\lambda )}\right] =\int
d^{3}x\,\varepsilon ^{\alpha \beta \gamma }V_{\alpha (\lambda )}\partial
_{\beta }A_{\gamma }^{\;\;(\lambda )},  \label{bf33}
\end{equation}
\begin{equation}
\delta _{\epsilon }V_{\alpha (\lambda )}=\partial _{\alpha }\epsilon
_{(\lambda )}^{\prime },\;\delta _{\epsilon }A_{\alpha }^{\;\;(\lambda
)}=\partial _{\alpha }\epsilon ^{(\lambda )},  \label{bf34}
\end{equation}
such that we deal with two types of Abelian covariant tensor gauge fields of
degree two.

The main scope of this paper is to analyze all consistent interactions that
can be added to action (\ref{bf33}), namely, all interactions that preserve
the field spectrum and the number of independent gauge transformations.
There are three main categories of such interactions: (I) the Lagrangian
action, the gauge transformations as well as their gauge algebra, are all
modified; (II) the action and the gauge symmetries are changed, but not
their algebra; (III) only the action is deformed, but not the gauge
symmetries. We want to deform the above free action by adding to it
interaction terms
\begin{equation}
I_{0}\rightarrow I=I_{0}+gI_{1}+g^{2}I_{2}+\cdots ,  \label{bf35}
\end{equation}
and meanwhile to modify the gauge transformations
\begin{equation}
\bar{\delta}_{\epsilon }V_{\alpha (\lambda )}=\partial _{\alpha }\epsilon
_{(\lambda )}^{\prime }+g\Theta _{\alpha \lambda }+O\left( g^{2}\right) ,\;%
\bar{\delta}_{\epsilon }A_{\alpha }^{\;\;(\lambda )}=\partial _{\alpha
}\epsilon ^{(\lambda )}+g\Delta _{\alpha }^{\;\;\lambda }+O\left(
g^{2}\right) ,  \label{bf36}
\end{equation}
such that the deformed action $I$ is invariant under the modified gauge
transformations at each order in the coupling constant $g$%
\begin{equation}
\bar{\delta}_{\epsilon }I=0.  \label{bf37}
\end{equation}
It is required that every term in the expansion (\ref{bf35}) be a local
functional in the corresponding fields. We focus on the determination of
non-trivial deformations, \textit{i.e.}, of those deformations of the action
that are not due to a local redefinition of the tensor gauge fields.

The most economical and elegant way to approach the consistent deformations
of a given gauge theory is to investigate the deformations of the rigid,
fermionic and nilpotent symmetry that encodes, among others, the entire
gauge structure of the theory--- the antibracket-antifield BRST symmetry. In
the case of the Abelian free model under discussion, the BRST differential $%
s $ reduces to the sum between the Koszul--Tate differential $\delta $ and
the exterior longitudinal derivative $\gamma $%
\begin{equation}
s=\delta +\gamma .  \label{bf38}
\end{equation}
While the Koszul--Tate differential is graded in terms of the antighost
number ($\mathrm{agh}$, $\mathrm{agh}\left( \delta \right) =-1$), the degree
associated with the exterior longitudinal derivative is named pure ghost
number ($\mathrm{pgh}$, $\mathrm{pgh}\left( \gamma \right) =1$). These
graduations do not interfere ($\mathrm{agh}\left( \gamma \right) =0=\mathrm{%
pgh}\left( \delta \right) $). The BRST differential is controlled by an
overall degree, called ghost number ($\mathrm{gh}$), and defined like the
difference between the pure ghost number and the antighost number, in terms
of which we have that $\mathrm{gh}\left( s\right) =\mathrm{gh}\left( \gamma
\right) =\mathrm{gh}\left( \delta \right) =1$. The second-order nilpotency
of $s$ ($s^{2}=0$) is equivalent to the nilpotency and anticommutation of
its components ($\delta ^{2}=0$, $\gamma ^{2}=0$, $\delta \gamma +\gamma
\delta =0$). The realization of these operators requires the introduction of
the fermionic ghosts $\left( \eta _{(\lambda )},C^{(\lambda )}\right) $
respectively corresponding to the gauge parameters $\left( \epsilon
_{(\lambda )}^{\prime },\epsilon ^{(\lambda )}\right) $, as well as of the
antifields ($V^{*\alpha (\lambda )},A_{\;\;\;(\lambda )}^{*\alpha },\eta
^{*(\lambda )},C_{(\lambda )}^{*}$), with Grassmann parity opposite to that
of the corresponding fields/ghosts. The pure ghost number and antighost
number of the BRST generators are valued like
\begin{equation}
\mathrm{pgh}\left( V_{\alpha (\lambda )}\right) =\mathrm{pgh}\left(
A_{\alpha }^{\;\;(\lambda )}\right) =0,\;\mathrm{pgh}\left( \eta _{(\lambda
)}\right) =\mathrm{pgh}\left( C^{(\lambda )}\right) =1,  \label{bf39}
\end{equation}
\begin{equation}
\mathrm{pgh}\left( V^{*\alpha (\lambda )}\right) =\mathrm{pgh}\left(
A_{\;\;\;(\lambda )}^{*\alpha }\right) =\mathrm{pgh}\left( \eta ^{*(\lambda
)}\right) =\mathrm{pgh}\left( C_{(\lambda )}^{*}\right) =0,  \label{bf40}
\end{equation}
\begin{equation}
\mathrm{agh}\left( V_{\alpha (\lambda )}\right) =\mathrm{agh}\left(
A_{\alpha }^{\;\;(\lambda )}\right) =\mathrm{agh}\left( \eta _{(\lambda
)}\right) =\mathrm{agh}\left( C^{(\lambda )}\right) =0,  \label{bf41}
\end{equation}
\begin{equation}
\mathrm{agh}\left( V^{*\alpha (\lambda )}\right) =\mathrm{agh}\left(
A_{\;\;\;(\lambda )}^{*\alpha }\right) =1,\;\mathrm{agh}\left( \eta
^{*(\lambda )}\right) =\mathrm{agh}\left( C_{(\lambda )}^{*}\right) =2,
\label{bf42}
\end{equation}
while the actions of $\delta $ and $\gamma $ on them read as
\begin{equation}
\gamma \left( V_{\alpha (\lambda )}\right) =\partial _{\alpha }\eta
_{(\lambda )},\;\gamma \left( A_{\alpha }^{\;\;(\lambda )}\right) =\partial
_{\alpha }C^{(\lambda )},\;\gamma \left( \eta _{(\lambda )}\right) =0=\gamma
\left( C^{(\lambda )}\right) ,  \label{bf43}
\end{equation}
\begin{equation}
\gamma \left( V^{*\alpha (\lambda )}\right) =\gamma \left( A_{\;\;\;(\lambda
)}^{*\alpha }\right) =\gamma \left( \eta ^{*(\lambda )}\right) =\gamma
\left( C_{(\lambda )}^{*}\right) =0,  \label{bf44}
\end{equation}
\begin{equation}
\delta \left( V_{\alpha (\lambda )}\right) =\delta \left( A_{\alpha
}^{\;\;(\lambda )}\right) =\delta \left( \eta _{(\lambda )}\right) =\delta
\left( C^{(\lambda )}\right) =0,  \label{bf45}
\end{equation}
\begin{equation}
\delta V^{*\alpha (\lambda )}=-\varepsilon ^{\alpha \beta \gamma }\partial
_{\beta }A_{\gamma }^{\;\;(\lambda )},\;\delta A_{\;\;\;(\lambda )}^{*\alpha
}=-\varepsilon ^{\alpha \beta \gamma }\partial _{\beta }V_{\gamma (\lambda
)},  \label{bf46}
\end{equation}
\begin{equation}
\delta \eta ^{*(\lambda )}=-\partial _{\alpha }V^{*\alpha (\lambda
)},\;\delta C_{(\lambda )}^{*}=-\partial _{\alpha }A_{\;\;\;(\lambda
)}^{*\alpha }.  \label{bf47}
\end{equation}
By construction, the Koszul--Tate differential realizes an homological
resolution of smooth functions defined on the stationary surface of the
field equations for $I_{0}$, $\delta I_{0}/\delta V_{\alpha (\lambda
)}\equiv \varepsilon ^{\alpha \beta \gamma }\partial _{\beta }A_{\gamma
}^{\;\;(\lambda )}=0$, $\delta I_{0}/\delta A_{\alpha }^{\;\;(\lambda
)}\equiv \varepsilon ^{\alpha \beta \gamma }\partial _{\beta }V_{\gamma
(\lambda )}=0$, while the cohomological space of the exterior longitudinal
derivative at pure ghost number zero computed in the homology of $\delta $
is nothing but the algebra of Lagrangian physical observables of the model
under consideration. Moreover, the cohomological space of the BRST
differential itself at ghost number zero $H^{0}\left( s\right) $, which
contains the so-called BRST observables, is isomorphic to the same algebra
of physical observables.

A remarkable feature of the BRST symmetry is that it is (anti)canonically
generated in a structure named antibracket, $s\cdot =\left( \cdot ,S\right) $%
, where the antibracket is obtained by decreeing the fields/ghosts
conjugated with the corresponding antifields. Its generator is bosonic, $%
\varepsilon \left( S\right) =0$, has ghost number zero, $\mathrm{gh}\left(
S\right) =0$, and satisfies the classical master equation
\begin{equation}
\left( S,S\right) =0,  \label{bf49}
\end{equation}
that expresses the nilpotency of $s$ at the canonical level. In our case the
solution to the master equation can be written like
\begin{equation}
S=I_{0}\left[ V_{\alpha (\lambda )},A_{\alpha }^{\;\;(\lambda )}\right]
+\int d^{3}x\left( V^{*\alpha (\lambda )}\partial _{\alpha }\eta _{(\lambda
)}+A_{\;\;\;(\lambda )}^{*\alpha }\partial _{\alpha }C^{(\lambda )}\right) .
\label{bf48}
\end{equation}
The solution to the classical master equation encodes all the information on
the gauge structure of $I_{0}$. Indeed, if we expand it according to the
antighost number, its antighost number zero component is precisely the
Lagrangian action $I_{0}$ of the gauge theory, while the antighost number
one terms describe the gauge symmetries of the action. The absence of
elements with antighost number higher than one is connected with the Abelian
and irreducible choice of the gauge generators.

\section{Deformation of the Free Model}

The reformulation of the problem of consistent deformations of action (\ref
{bf33}) and of its gauge symmetries (\ref{bf34}) in terms of the BRST
setting is based on the observation that if a deformation (\ref{bf35}--\ref
{bf36}) of the classical theory can be consistently constructed, then the
solution (\ref{bf48}) to the master equation (\ref{bf49}) for the free
theory can be deformed into
\begin{equation}
\bar{S}=S+gS_{1}+g^{2}S_{2}+O\left( g^{3}\right) ,\;\varepsilon \left( \bar{S%
}\right) =0,\;\mathrm{gh}\left( \bar{S}\right) =0,  \label{bf50}
\end{equation}
such that
\begin{equation}
\left( \bar{S},\bar{S}\right) =0.  \label{bf51}
\end{equation}
The projection of (\ref{bf51}) along the various powers in the coupling
constant induces the following tower of equations:
\begin{equation}
g^{0}:\left( S,S\right) =0,  \label{bf52}
\end{equation}
\begin{equation}
g^{1}:\left( S_{1},S\right) =0,  \label{bf53}
\end{equation}
\begin{equation}
g^{2}:\frac{1}{2}\left( S_{1},S_{1}\right) +\left( S_{2},S\right) =0,
\label{bf54}
\end{equation}
\[
\vdots
\]
The first equation is satisfied by hypothesis. The second one governs the
first-order deformation of the solution to the master equation, $S_{1}$, and
it expresses the fact that $S_{1}$ is a BRST co-cycle, $sS_{1}=0$, and hence
it exists and is local. The remaining equations are responsible for the
higher-order deformations of the solution to the master equation. No
obstructions arise in finding solutions to them as long as no further
restrictions, such as space-time locality, are imposed. Taking into account (%
\ref{bf35}--\ref{bf36}), the first two pieces of $S_{1}$ regarded as an
expansion according to the antighost number
\begin{equation}
S_{1}=\stackrel{[0]}{S}_{1}+\stackrel{[1]}{S}_{1}+\cdots \stackrel{[j]}{S}%
_{1},\;\varepsilon \left( \stackrel{[k]}{S}_{1}\right) =0,\;\mathrm{gh}%
\left( \stackrel{[k]}{S}_{1}\right) =0,\;\mathrm{agh}\left( \stackrel{[k]}{S}%
_{1}\right) =k,  \label{bf55}
\end{equation}
are exactly
\begin{equation}
\stackrel{\lbrack 0]}{S}_{1}=I_{1},\;\stackrel{[1]}{S}_{1}=\int d^{3}x\left(
V^{*\alpha (\lambda )}\bar{\Theta}_{\alpha \lambda }+A_{\;\;\;(\lambda
)}^{*\alpha }\bar{\Delta}_{\alpha }^{\;\;\lambda }\right) ,  \label{bf56}
\end{equation}
where $\bar{\Theta}_{\alpha \lambda }$ and $\bar{\Delta}_{\alpha
}^{\;\;\lambda }$ are obtained from $\Theta _{\alpha \lambda }$ and $\Delta
_{\alpha }^{\;\;\lambda }$ in (\ref{bf36}) where we replace the gauge
parameters by the ghosts. Obviously, only non-trivial first-order
deformations should be considered, since trivial ones ($S_{1}=sB$) lead to
trivial deformations of the initial theory, and can be eliminated by
convenient redefinitions of the fields. Ignoring the trivial deformations,
it follows that $S_{1}$ is a non-trivial BRST-observable, $S_{1}\in
H^{0}\left( s\right) $. Once that the deformation Eqs.~(\ref{bf53}--\ref
{bf54}), etc., have been solved by means of specific cohomological
techniques, from the consistent non-trivial deformed solution to the master
equation we can extract all the information on the gauge structure of the
accompanying interacting theory. Eventually, we can make a selection of the
coupled models with special physical characteristics, like space-time
locality, PT-invariance, etc.

Although the cohomological approach to the deformation of gauge theories
seems to deal with the same equations as the standard one, it is far more
effective as it organizes the recursive construction in a systematic manner,
precisely relates the consistent interactions to co-cycles of the BRST
differential, reduces the trivial deformations to trivial classes of the
BRST cohomology, and brings in the powerful tools of homological algebra.

\subsection{First-order deformation}

Initially, we approach the Eq.~(\ref{bf53}), responsible for the first-order
deformation of the solution to the master equation. By making the notation $%
S_{1}=\int d^{3}x\,a_{1}$, this equation takes the local form
\begin{equation}
sa_{1}=\partial _{\mu }j^{\mu },  \label{bf57}
\end{equation}
for some local current $j^{\mu }$. The established fact that $S_{1}$ belongs
to $H^{0}\left( s\right) $ implies that $a_{1}$ pertains to the zeroth order
local cohomological space of the BRST differential, $H^{0}\left( s|d\right) $%
, where $d$ is the exterior space-time differential. Then, the trivial
solutions to (\ref{bf57}), namely, $a_{1}=sb+\partial _{\mu }k^{\mu }$, can
be discarded. In order to solve the equation (\ref{bf57}), we recall the
expansion (\ref{bf55}), where we make the notations $\stackrel{[k]}{S}%
_{1}=\int d^{3}x\stackrel{[k]}{a}_{1}$, which provides the development of $%
a_{1}$ along the antighost number
\begin{equation}
a_{1}=\stackrel{[0]}{a}_{1}+\stackrel{[1]}{a}_{1}+\cdots +\stackrel{[j]}{a}%
_{1},\;\varepsilon \left( \stackrel{[k]}{a}_{1}\right) =0,\;\mathrm{gh}%
\left( \stackrel{[k]}{a}_{1}\right) =0,\;\mathrm{agh}\left( \stackrel{[k]}{a}%
_{1}\right) =k.  \label{bf58}
\end{equation}
The number of terms in (\ref{bf58}) is finite and it can be proven that the
last representative can be taken to be annihilated by the exterior
longitudinal derivative, $\gamma \stackrel{[j]}{a}_{1}=0$, with the
precaution that it should not be trivial ($\gamma $-exact), $\stackrel{[j]}{a%
}_{1}\neq \gamma c$. Then, as $\stackrel{[j]}{a}_{1}$ belongs to $H\left(
\gamma \right) $, we have to compute the cohomology of $\gamma $ in order to
generate the component of highest antighost number from the first-order
deformation. Looking at the definitions (\ref{bf43}--\ref{bf44}), we read
that $H\left( \gamma \right) $ is spanned by the polynomials in the
(undifferentiated) ghosts with coefficients that are functions of the `field
strengths' $\partial _{\left[ \alpha \right. }V_{\left. \beta \right]
(\lambda )}$, $\partial _{\left[ \alpha \right. }A_{\left. \beta \right]
}^{\;\;(\lambda )}$, of the antifields $\chi ^{*}\equiv \left( V^{*\alpha
(\lambda )},A_{\;\;\;(\lambda )}^{*\alpha },\eta ^{*(\lambda )},C_{(\lambda
)}^{*}\right) $, as well as of their space-time derivatives up to a finite
order. The space-time derivatives of the ghosts are removed from $H\left(
\gamma \right) $ since their first-order derivatives are already $\gamma $%
-exact, as can be observed from the first and second definitions in (\ref
{bf43}). In consequence, we have that
\begin{equation}
\stackrel{\lbrack j]}{a}_{1}=\alpha _{j}\left( \left[ \partial _{\left[
\alpha \right. }V_{\left. \beta \right] (\lambda )}\right] ,\left[ \partial
_{\left[ \alpha \right. }A_{\left. \beta \right] }^{\;\;(\lambda )}\right]
,\left[ \chi ^{*}\right] \right) e^{j}\left( \eta _{(\lambda )},C^{(\lambda
)}\right) ,  \label{bf59}
\end{equation}
where $e^{j}$ are the elements of pure ghost number $j$ of a basis in the
ghosts, the notation $f\left( \left[ q\right] \right) $ signifies that $f$
depends on $q$ and its space-time derivatives up to a finite order, and the
coefficients have $\mathrm{agh}\left( \alpha _{j}\right) =j$. Inserting (\ref
{bf59}) into the Eq.~(\ref{bf57}) projected on antighost number $\left(
j-1\right) $%
\begin{equation}
\delta \stackrel{[j]}{a}_{1}+\gamma \stackrel{[j-1]}{a}_{1}=\partial _{\mu
}m^{\mu },  \label{bf60}
\end{equation}
we obtain that a necessary condition for the existence of $\stackrel{[j-1]}{a%
}_{1}$ is that the coefficients $\alpha _{j}$ belong to the local homology
of the Koszul--Tate differential at antighost number $j$, $\alpha _{j}\in
H_{j}\left( \delta |d\right) $, \textit{i.e.}, $\delta \alpha _{j}=\partial
_{\mu }d^{\mu }$. The free model under investigation is described by a
normal gauge theory of Cauchy order two, and thus we have that the local
homology of the Koszul--Tate differential vanishes~\cite{gen1} at antighost
numbers strictly greater than two, $H_{i}\left( \delta |d\right) =0$ for $i>2
$. This forces the development (\ref{bf58}) to stop at antighost number two
\begin{equation}
a_{1}=\stackrel{[0]}{a}_{1}+\stackrel{[1]}{a}_{1}+\stackrel{[2]}{a}_{1},
\label{bf61}
\end{equation}
where $\gamma \stackrel{[2]}{a}_{1}=0$, such that $\stackrel{[2]}{a}_{1}$ is
of the form (\ref{bf59}), with $j=2$ and $\alpha _{2}$ from $H_{2}\left(
\delta |d\right) $. On the one hand, the most general representatives of $%
H_{2}\left( \delta |d\right) $ are
\begin{equation}
k_{1}\eta ^{*(\lambda )}+k_{2}C_{(\lambda )}^{*},  \label{bf62}
\end{equation}
with $k_{1,2}$ real constants, and, on the other hand, the elements of pure
ghost number two of a basis in the ghosts are
\begin{equation}
\eta _{(\mu )}\eta _{(\nu )},\eta _{(\mu )}C^{(\nu )},C^{(\mu )}C^{(\nu )},
\label{bf63}
\end{equation}
such that $\stackrel{[2]}{a}_{1}$ is of the form
\begin{eqnarray}
&&\stackrel{\lbrack 2]}{a}_{1}=\eta _{(\lambda )}^{*}\left( f_{1}^{\lambda
\mu \nu }\eta _{(\mu )}\eta _{(\nu )}+f_{2}^{\lambda \mu \nu }\eta _{(\mu
)}C_{(\nu )}+f_{3}^{\lambda \mu \nu }C_{(\mu )}C_{(\nu )}\right)  \nonumber
\\
&&+C_{(\lambda )}^{*}\left( f_{4}^{\lambda \mu \nu }\eta _{(\mu )}\eta
_{(\nu )}+f_{5}^{\lambda \mu \nu }\eta _{(\mu )}C_{(\nu )}+f_{6}^{\lambda
\mu \nu }C_{(\mu )}C_{(\nu )}\right) ,  \label{bf64}
\end{eqnarray}
where $\left( f_{a}^{\lambda \mu \nu }\right) _{a=\overline{1,6}}$ must be
non-derivative constants. (If any of these constants contains at least one
space-time derivative, then the corresponding term in $\stackrel{[2]}{a}_{1}$
is $\gamma $-exact, and thus trivial in $H\left( \gamma \right) $.) By
covariance arguments, all these constants can only be proportional with the
completely antisymmetric symbol in three dimensions, such that the last term
in the first-order deformation (\ref{bf61}) becomes (up to trivial $\gamma $%
-exact solutions)
\begin{eqnarray}
&&\stackrel{\lbrack 2]}{a}_{1}=\varepsilon ^{\lambda \mu \nu }\left( \eta
_{(\lambda )}^{*}\left( c_{1}\eta _{(\mu )}\eta _{(\nu )}+c_{2}\eta _{(\mu
)}C_{(\nu )}+c_{3}C_{(\mu )}C_{(\nu )}\right) \right.  \nonumber \\
&&\left. +C_{(\lambda )}^{*}\left( c_{4}\eta _{(\mu )}\eta _{(\nu
)}+c_{5}\eta _{(\mu )}C_{(\nu )}+c_{6}C_{(\mu )}C_{(\nu )}\right) \right) ,
\label{bf65}
\end{eqnarray}
with $c_{1}$, etc., arbitrary real constants.

Taking into account that the equation for $\stackrel{[1]}{a}_{1}$ is
obtained from (\ref{bf60}) where we set $j=2$%
\begin{equation}
\delta \stackrel{[2]}{a}_{1}+\gamma \stackrel{[1]}{a}_{1}=\partial _{\mu
}m^{\mu },  \label{bf66}
\end{equation}
with the help of (\ref{bf65}) and (\ref{bf43}--\ref{bf47}) we arrive at
\begin{eqnarray}
&&\stackrel{\lbrack 1]}{a}_{1}=\varepsilon ^{\lambda \mu \nu }\left(
V_{\;\;\;(\lambda )}^{*\alpha }\left( \left( 2c_{1}V_{\alpha (\mu
)}+c_{2}A_{\alpha (\mu )}\right) \eta _{(\nu )}+\left( 2c_{3}A_{\alpha (\mu
)}+c_{2}V_{\alpha (\mu )}\right) C_{(\nu )}\right) +\right.  \nonumber \\
&&\left. A_{\;\;\;(\lambda )}^{*\alpha }\left( \left( 2c_{6}A_{\alpha (\mu
)}+c_{5}V_{\alpha (\mu )}\right) C_{(\nu )}+\left( 2c_{4}V_{\alpha (\mu
)}+c_{5}A_{\alpha (\mu )}\right) \eta _{(\nu )}\right) \right) +\stackrel{[1]%
}{a^{\prime }}_{1},  \label{bf67}
\end{eqnarray}
where $\stackrel{[1]}{a^{\prime }}_{1}$ denotes the general non-trivial
solution to the `homogeneous' equation $\gamma \stackrel{[1]}{a^{\prime }}%
_{1}=\partial _{\mu }m^{\prime \mu }$. Such solutions stem from $\stackrel{%
[2]}{a^{\prime }}_{1}=0$, such that they do not modify the gauge algebra,
but deform the gauge transformations (type (II) interactions). Following a
standard reasoning, it can be shown that the above `homogeneous' equation is
equivalent, up to trivial redefinitions, to the equation with vanishing $%
m^{\prime \mu }$%
\begin{equation}
\gamma \stackrel{[1]}{a^{\prime }}_{1}=0.  \label{bf68}
\end{equation}
According to (\ref{bf59}), the general solution to the Eq.~(\ref{bf68}) is
of the type
\begin{eqnarray}
&&\stackrel{\lbrack 1]}{a^{\prime }}_{1}=\alpha _{1}^{\lambda }\left( \left[
\partial _{\left[ \alpha \right. }V_{\left. \beta \right] (\lambda )}\right]
,\left[ \partial _{\left[ \alpha \right. }A_{\left. \beta \right]
}^{\;\;(\lambda )}\right] ,\left[ V^{*\alpha (\lambda )}\right] ,\left[
A_{\;\;\;(\lambda )}^{*\alpha }\right] \right) \eta _{(\lambda )}  \nonumber
\\
&&+\alpha _{1}^{\prime \lambda }\left( \left[ \partial _{\left[ \alpha
\right. }V_{\left. \beta \right] (\lambda )}\right] ,\left[ \partial
_{\left[ \alpha \right. }A_{\left. \beta \right] }^{\;\;(\lambda )}\right]
,\left[ V^{*\alpha (\lambda )}\right] ,\left[ A_{\;\;\;(\lambda )}^{*\alpha
}\right] \right) C_{(\lambda )},  \label{bf69}
\end{eqnarray}
since the elements with pure ghost number equal to one of a basis in the
ghosts are precisely $\eta _{(\lambda )}$ and $C_{(\lambda )}$. (The
antifields $\eta ^{*(\lambda )}$, $C_{(\lambda )}^{*}$ and their space-time
derivatives are not allowed to enter $\stackrel{[1]}{a^{\prime }}_{1}$ as
they have the antighost number equal to two.) Meanwhile, the quantities $%
\partial _{\left[ \alpha \right. }V_{\left. \beta \right] (\lambda )}$, $%
\partial _{\left[ \alpha \right. }A_{\left. \beta \right] }^{\;\;(\lambda )}$
and their space-time derivatives are already $\delta $-exact (see (\ref{bf46}%
)), such that the dependence of $\alpha _{1}^{\lambda }$ and $\alpha
_{1}^{\prime \lambda }$ on them leads to trivial ($\delta $-exact and $%
\gamma $-closed, hence $s$-exact) terms, which further implies that
\begin{equation}
\alpha _{1}^{\lambda }=\alpha _{1}^{\lambda }\left( \left[ V^{*\alpha
(\lambda )}\right] ,\left[ A_{\;\;\;(\lambda )}^{*\alpha }\right] \right)
,\;\alpha _{1}^{\prime \lambda }=\alpha _{1}^{\prime \lambda }\left( \left[
V^{*\alpha (\lambda )}\right] ,\left[ A_{\;\;\;(\lambda )}^{*\alpha }\right]
\right) .  \label{bf70}
\end{equation}
On the other hand, $\alpha _{1}^{\lambda }$ and $\alpha _{1}^{\prime \lambda
}$ can only be linear in the undifferentiated antifields $V^{*\alpha
(\lambda )}$ and $A_{\;\;\;(\lambda )}^{*\alpha }$. This is because the
appearance of at least one space-time derivative in these antifields leads,
via an integration by parts in the corresponding functional $\stackrel{[1]}{%
S^{\prime }}_{1}=\int d^{3}\stackrel{[1]}{a^{\prime }}_{1}$, to a $\gamma $%
-exact (and hence trivial) solution to (\ref{bf68}), which can always be
taken to vanish. In consequence, the most general solution to the
`homogeneous' Eq.~(\ref{bf68}) is expressed (up to trivial $\gamma $-exact
solutions) by
\begin{equation}
\stackrel{\lbrack 1]}{a^{\prime }}_{1}=\varepsilon ^{\alpha \lambda \mu
}\left( V_{\alpha (\lambda )}^{*}\left( c_{1}^{\prime }C_{(\mu
)}+c_{2}^{\prime }\eta _{(\mu )}\right) +A_{\alpha (\lambda )}^{*}\left(
c_{3}^{\prime }C_{(\mu )}+c_{4}^{\prime }\eta _{(\mu )}\right) \right) ,
\label{bf71}
\end{equation}
with $c_{1}^{\prime }$ and so on arbitrary real constants.

By projecting (\ref{bf57}) on antighost number zero, we deduce the equation
verified by the antighost number zero component of the first-order
deformation
\begin{equation}
\delta \stackrel{[1]}{a}_{1}+\gamma \stackrel{[0]}{a}_{1}=\partial _{\mu
}n^{\mu }.  \label{bf72}
\end{equation}
Using the expression (\ref{bf67}) of $\stackrel{[1]}{a}_{1}$ with $\stackrel{%
[1]}{a^{\prime }}_{1}$ given by (\ref{bf71}), by direct computation we infer
that
\begin{eqnarray}
&&\delta \stackrel{[1]}{a}_{1}=\partial _{\mu }n^{\prime \mu }+\varepsilon
^{\lambda \mu \nu }\varepsilon ^{\alpha \beta \gamma }\left( \left(
2c_{1}\left( \partial _{\gamma }A_{\alpha (\lambda )}\right) V_{\beta (\mu
)}+c_{5}A_{\alpha (\lambda )}\partial _{\gamma }V_{\beta (\mu )}\right) \eta
_{(\nu )}\right.  \nonumber \\
&&+\left( 2c_{6}\left( \partial _{\gamma }V_{\alpha (\lambda )}\right)
A_{\beta (\mu )}+c_{2}V_{\alpha (\lambda )}\partial _{\gamma }A_{\beta (\mu
)}\right) C_{(\nu )}  \nonumber \\
&&\left. -\frac{1}{2}\left( c_{5}V_{\beta (\mu )}V_{\gamma (\nu )}\gamma
\left( A_{\alpha (\lambda )}\right) +c_{2}A_{\beta (\mu )}A_{\gamma (\nu
)}\gamma \left( V_{\alpha (\lambda )}\right) \right) \right)  \nonumber \\
&&+c_{2}^{\prime }\left( A_{\;\;(\beta )}^{\beta }\gamma \left(
V_{\;\;(\gamma )}^{\gamma }\right) -A_{\;\;(\beta )}^{\gamma }\gamma \left(
V_{\;\;(\gamma )}^{\beta }\right) \right)  \nonumber \\
&&+c_{3}^{\prime }\left( V_{\;\;(\gamma )}^{\gamma }\gamma \left(
A_{\;\;(\beta )}^{\beta }\right) -V_{\;\;(\gamma )}^{\beta }\gamma \left(
A_{\;\;(\beta )}^{\gamma }\right) \right)  \nonumber \\
&&-\gamma \left( \frac{1}{3}\varepsilon ^{\lambda \mu \nu }\varepsilon
^{\alpha \beta \gamma }\left( c_{3}A_{\alpha (\lambda )}A_{\beta (\mu
)}A_{\gamma (\nu )}+c_{4}V_{\alpha (\lambda )}V_{\beta (\mu )}V_{\gamma (\nu
)}\right) \right.  \nonumber \\
&&+\frac{c_{1}^{\prime }}{2}\left( A_{\;\;(\beta )}^{\gamma }A_{\;\;(\gamma
)}^{\beta }-A_{\;\;(\beta )}^{\beta }A_{\;\;(\gamma )}^{\gamma }\right)
\nonumber \\
&&\left. +\frac{c_{4}^{\prime }}{2}\left( V_{\;\;(\beta )}^{\gamma
}V_{\;\;(\gamma )}^{\beta }-V_{\;\;(\beta )}^{\beta }V_{\;\;(\gamma
)}^{\gamma }\right) \right) ,  \label{bf73}
\end{eqnarray}
such that the existence of $\stackrel{[0]}{a}_{1}$ implies the following
conditions on the various constants
\begin{equation}
2c_{1}=c_{5},\;2c_{6}=c_{2},\;c_{2}^{\prime }=c_{3}^{\prime }.  \label{bf74}
\end{equation}
Replacing (\ref{bf74}) in (\ref{bf73}), the solution to the Eq.~(\ref{bf72}%
), expressed in terms of seven independent real constants, reads as
\begin{eqnarray}
&&\stackrel{\lbrack 0]}{a}_{1}=\varepsilon ^{\lambda \mu \nu }\varepsilon
^{\alpha \beta \gamma }\left( \frac{1}{2}\left( c_{2}V_{\alpha (\lambda
)}A_{\beta (\mu )}A_{\gamma (\nu )}+c_{5}A_{\alpha (\lambda )}V_{\beta (\mu
)}V_{\gamma (\nu )}\right) \right.  \nonumber \\
&&\left. +\frac{1}{3}\left( c_{3}A_{\alpha (\lambda )}A_{\beta (\mu
)}A_{\gamma (\nu )}+c_{4}V_{\alpha (\lambda )}V_{\beta (\mu )}V_{\gamma (\nu
)}\right) \right)  \nonumber \\
&&+\frac{1}{2}\left( c_{1}^{\prime }\left( A_{\;\;(\beta )}^{\gamma
}A_{\;\;(\gamma )}^{\beta }-A_{\;\;(\beta )}^{\beta }A_{\;\;(\gamma
)}^{\gamma }\right) +c_{4}^{\prime }\left( V_{\;\;(\beta )}^{\gamma
}V_{\;\;(\gamma )}^{\beta }-V_{\;\;(\beta )}^{\beta }V_{\;\;(\gamma
)}^{\gamma }\right) \right)  \nonumber \\
&&+c_{2}^{\prime }\left( A_{\;\;(\beta )}^{\gamma }V_{\;\;(\gamma )}^{\beta
}-A_{\;\;(\beta )}^{\beta }V_{\;\;(\gamma )}^{\gamma }\right) +\stackrel{[0]%
}{a^{\prime \prime }}_{1}\equiv \stackrel{[0]}{\bar{a}}_{1}+\stackrel{[0]}{%
a^{\prime \prime }}_{1}.  \label{bf75}
\end{eqnarray}
In the above, $\stackrel{[0]}{a^{\prime \prime }}_{1}$ represents the
general non-trivial solution to the `homogeneous' equation
\begin{equation}
\gamma \stackrel{[0]}{a^{\prime \prime }}_{1}=\partial _{\mu }n^{\prime
\prime \mu }.  \label{bf76}
\end{equation}
Such solutions come from $\stackrel{[1]}{a^{\prime \prime }}_{1}=0$, such
that they do not deform either the gauge algebra or the gauge
transformations (type (III) interactions), but simply add to the original
action $\gamma $-invariant modulo $d$ terms. There are two types of
solutions to (\ref{bf76}). The first one corresponds to $n^{\prime \prime
\mu }=0$ and is given by arbitrary polynomials in $\partial _{\left[ \alpha
\right. }V_{\left. \beta \right] (\lambda )}$, $\partial _{\left[ \alpha
\right. }A_{\left. \beta \right] }^{\;\;(\lambda )}$ and their space-time
derivatives. However, this type of solutions is easily seen to be trivial ($%
s $-exact) due to the simultaneous $\delta $-exactness and $\gamma $%
-closedness of $\partial _{\left[ \alpha \right. }V_{\left. \beta \right]
(\lambda )}$ and $\partial _{\left[ \alpha \right. }A_{\left. \beta \right]
}^{\;\;(\lambda )}$, and hence they can be removed by taking the associated $%
\stackrel{[0]}{a^{\prime \prime }}_{1}$ to vanish. The second one leads to
non-vanishing local currents $n^{\prime \prime \mu }$ and is expressed by
generalized Chern--Simons terms
\begin{eqnarray}
&&\stackrel{\lbrack 0]}{a^{\prime \prime }}_{1}=\varepsilon ^{\alpha \beta
\gamma }\left( \sum_{k=0}^{m}d_{k}\left( \partial _{\mu _{1}\cdots \mu
_{k}}A_{\alpha (\lambda )}\right) \left( \partial ^{\mu _{1}\cdots \mu
_{k}}\partial _{\beta }A_{\gamma }^{\;\;(\lambda )}\right) \right.  \nonumber
\\
&&+\sum_{k=0}^{m^{\prime }}d_{k}^{\prime }\left( \partial _{\mu _{1}\cdots
\mu _{k}}V_{\alpha (\lambda )}\right) \left( \partial ^{\mu _{1}\cdots \mu
_{k}}\partial _{\beta }V_{\gamma }^{\;\;(\lambda )}\right)  \nonumber \\
&&\left. +\sum_{k=0}^{m^{\prime \prime }}d_{k}^{\prime \prime }\left(
\partial _{\mu _{1}\cdots \mu _{k}}V_{\alpha (\lambda )}\right) \left(
\partial ^{\mu _{1}\cdots \mu _{k}}\partial _{\beta }A_{\gamma
}^{\;\;(\lambda )}\right) \right) ,  \label{bf77}
\end{eqnarray}
where all $d_{k}$, $d_{k}^{\prime }$ and $d_{k}^{\prime \prime }$ are
constants. The solution (\ref{bf77}) leads to a deformed Lagrangian action
with fields equations involving more than one derivative in the fields,
unlike the original action, which has first-order derivative field
equations. In order to preserve the first-order character at the level of
the deformed field equations, we are tempted to forbid this discontinuous
behavior by dropping out all the terms from (\ref{bf77}) with $k>0$. We show
that this is not necessary as $\stackrel{[0]}{a^{\prime \prime }}_{1}$ like
in (\ref{bf77}) can be removed from $a_{1}$ by adding to it trivial terms.
Indeed, it is easy to see that $\stackrel{[0]}{a^{\prime \prime }}_{1} $ is $%
\delta $-exact
\begin{eqnarray}
&&\stackrel{\lbrack 0]}{a^{\prime \prime }}_{1}=-\delta \left(
\sum_{k=0}^{m}d_{k}\left( \partial ^{\mu _{1}\cdots \mu _{k}}V^{*\alpha
(\lambda )}\right) \left( \partial _{\mu _{1}\cdots \mu _{k}}A_{\alpha
(\lambda )}\right) \right.  \nonumber \\
&&+\sum_{k=0}^{m^{\prime }}d_{k}^{\prime }\left( \partial ^{\mu _{1}\cdots
\mu _{k}}A^{*\alpha (\lambda )}\right) \left( \partial _{\mu _{1}\cdots \mu
_{k}}V_{\alpha (\lambda )}\right)  \nonumber \\
&&\left. +\sum_{k=0}^{m^{\prime \prime }}d_{k}^{\prime \prime }\left(
\partial ^{\mu _{1}\cdots \mu _{k}}V^{*\alpha (\lambda )}\right) \left(
\partial _{\mu _{1}\cdots \mu _{k}}V_{\alpha (\lambda )}\right) \right)
\equiv \delta T.  \label{bf78a}
\end{eqnarray}
In the meantime, the solution (\ref{bf71}) to the `homogeneous' Eq.~(\ref
{bf68}) is unique up to a trivial ($\gamma $-exact) quantity, so we can
replace it by
\begin{equation}
\stackrel{\lbrack 1]}{a^{\prime }}_{1}\rightarrow \stackrel{[1]}{\bar{a}%
^{\prime }}_{1}=\stackrel{[1]}{a^{\prime }}_{1}+\gamma T.  \label{bf78b}
\end{equation}
In agreement with (\ref{bf76}) and (\ref{bf78a}), this trivial term simply
adds a divergence to the $\delta $-variation of the corresponding $\stackrel{%
[1]}{a}_{1}$
\begin{equation}
\stackrel{\lbrack 1]}{a}_{1}\rightarrow \stackrel{[1]}{\bar{a}}_{1}=%
\stackrel{[1]}{a}_{1}+\gamma T,\;\delta \stackrel{[1]}{a}_{1}\rightarrow
\delta \stackrel{[1]}{\bar{a}}_{1}=\delta \stackrel{[1]}{a}_{1}+\partial
_{\mu }n^{\prime \prime \mu },  \label{bf78f}
\end{equation}
such that (\ref{bf75}) remains the general solution also to the Eq.~(\ref
{bf72}) with $\stackrel{[1]}{\bar{a}}_{1}$ instead of $\stackrel{[1]}{a}_{1}$%
. Then, according to (\ref{bf38}), (\ref{bf75}), (\ref{bf78a}) and (\ref
{bf78f}), we have that the introduction of this term amounts in the
first-order deformation to
\begin{equation}
a_{1}\rightarrow \bar{a}_{1}=\stackrel{[2]}{a}_{1}+\stackrel{[1]}{\bar{a}}%
_{1}+\stackrel{[0]}{a}_{1}=\stackrel{[2]}{a}_{1}+\stackrel{[1]}{a}_{1}+%
\stackrel{[0]}{\bar{a}}_{1}+sT.  \label{bf78c}
\end{equation}
As the first-order deformation is unique up to trivial ($s$-exact modulo $d$%
) elements, we can safely discard $sT$ from $\bar{a}_{1}$, which is
equivalent to setting
\begin{equation}
\stackrel{\lbrack 0]}{a^{\prime \prime }}_{1}=0,  \label{eq}
\end{equation}
in (\ref{bf75}). In consequence, we will work with the non-trivial
first-order deformation of the solution to the master equation
\begin{equation}
a_{1}=\stackrel{[2]}{a}_{1}+\stackrel{[1]}{a}_{1}+\stackrel{[0]}{\bar{a}}%
_{1}.  \label{bf78d}
\end{equation}
We cannot stress enough that no assumption on the number of derivatives in $%
a_{1}$ is necessary, since all higher-order derivative terms can be
completely eliminated from $\stackrel{[0]}{a}_{1}$ by means of trivial
quantities.

So far, we obtained the most general expression of the non-trivial
first-order deformation of the solution to the master equation for the model
under study like in (\ref{bf78d}), where its various components are pictured
by (\ref{bf65}), (\ref{bf67}), (\ref{bf71}), (\ref{bf75}) and (\ref{eq}),
and the constants involved are subject to the conditions (\ref{bf74}).
Putting together all these results, we arrive at
\begin{eqnarray}
&&S_{1}=\int d^{3}x\left( \varepsilon ^{\lambda \mu \nu }\left( \eta
_{(\lambda )}^{*}\left( \frac{c_{5}}{2}\eta _{(\mu )}\eta _{(\nu
)}+c_{2}\eta _{(\mu )}C_{(\nu )}+c_{3}C_{(\mu )}C_{(\nu )}\right) \right.
\right.  \nonumber \\
&&+C_{(\lambda )}^{*}\left( \frac{c_{2}}{2}C_{(\mu )}C_{(\nu )}+c_{5}C_{(\mu
)}\eta _{(\nu )}+c_{4}\eta _{(\mu )}\eta _{(\nu )}\right)  \nonumber \\
&&+V_{\;\;\;(\lambda )}^{*\alpha }\left( \left( c_{5}V_{\alpha (\mu
)}+c_{2}A_{\alpha (\mu )}\right) \eta _{(\nu )}+\left( 2c_{3}A_{\alpha (\mu
)}+c_{2}V_{\alpha (\mu )}\right) C_{(\nu )}\right)  \nonumber \\
&&+A_{\;\;\;(\lambda )}^{*\alpha }\left( \left( c_{2}A_{\alpha (\mu
)}+c_{5}V_{\alpha (\mu )}\right) C_{(\nu )}+\left( 2c_{4}V_{\alpha (\mu
)}+c_{5}A_{\alpha (\mu )}\right) \eta _{(\nu )}\right)  \nonumber \\
&&+\varepsilon ^{\alpha \beta \gamma }\left( \frac{1}{2}\left(
c_{2}V_{\alpha (\lambda )}A_{\beta (\mu )}A_{\gamma (\nu )}+c_{5}A_{\alpha
(\lambda )}V_{\beta (\mu )}V_{\gamma (\nu )}\right) \right.  \nonumber \\
&&\left. \left. +\frac{1}{3}\left( c_{3}A_{\alpha (\lambda )}A_{\beta (\mu
)}A_{\gamma (\nu )}+c_{4}V_{\alpha (\lambda )}V_{\beta (\mu )}V_{\gamma (\nu
)}\right) \right) \right)  \nonumber \\
&&+\varepsilon ^{\alpha \lambda \mu }\left( V_{\alpha (\lambda )}^{*}\left(
c_{1}^{\prime }C_{(\mu )}+c_{2}^{\prime }\eta _{(\mu )}\right) +A_{\alpha
(\lambda )}^{*}\left( c_{4}^{\prime }\eta _{(\mu )}+c_{2}^{\prime }C_{(\mu
)}\right) \right)  \nonumber \\
&&+\frac{c_{1}^{\prime }}{2}\left( A_{\;\;(\beta )}^{\gamma }A_{\;\;(\gamma
)}^{\beta }-A_{\;\;(\beta )}^{\beta }A_{\;\;(\gamma )}^{\gamma }\right) +%
\frac{c_{4}^{\prime }}{2}\left( V_{\;\;(\beta )}^{\gamma }V_{\;\;(\gamma
)}^{\beta }-V_{\;\;(\beta )}^{\beta }V_{\;\;(\gamma )}^{\gamma }\right)
\nonumber \\
&&\left. +c_{2}^{\prime }\left( A_{\;\;(\beta )}^{\gamma }V_{\;\;(\gamma
)}^{\beta }-A_{\;\;(\beta )}^{\beta }V_{\;\;(\gamma )}^{\gamma }\right)
\right) .  \label{bf79}
\end{eqnarray}
It has a beautiful symmetric form with respect to the permutation of the two
tensors, accompanying BRST generators and arbitrary constants, being
invariant under the transformation
\begin{equation}
V_{\alpha (\lambda )}\longleftrightarrow A_{\alpha (\lambda )},\;V^{*\alpha
(\lambda )}\longleftrightarrow A^{*\alpha (\lambda )},\;\eta _{(\nu
)}\longleftrightarrow C_{(\nu )},\;\eta ^{*(\nu )}\longleftrightarrow
C^{*(\nu )},  \label{bf81}
\end{equation}
\begin{equation}
c_{2}\longleftrightarrow c_{5},\;c_{3}\longleftrightarrow
c_{4},\;c_{1}^{\prime }\longleftrightarrow c_{4}^{\prime },\;c_{2}^{\prime
}\longleftrightarrow c_{2}^{\prime }.  \label{bf80}
\end{equation}

\subsection{Higher-order deformations}

The next step of the deformation procedure consists in the determination of
the second- and higher-order deformations of the solution to the master
equation. The second-order deformation is subject to the Eq.~(\ref{bf54}),
and it shows that the existence of local solutions $S_{2}$ requires that the
antibracket $\left( S_{1},S_{1}\right) $ is the $s$-variation of a local
functional. By direct computation, we find that
\begin{eqnarray}
&&\frac{1}{2}\left( S_{1},S_{1}\right) =\int d^{3}x\left( \left(
4c_{3}c_{4}-c_{2}c_{5}\right) \left( \eta ^{*(\lambda )}C^{(\mu )}\eta
_{(\lambda )}\eta _{(\mu )}+C^{*(\lambda )}\eta ^{(\mu )}C_{(\lambda
)}C_{(\mu )}\right. \right.  \nonumber \\
&&+V^{*\alpha (\lambda )}\left( \left( V_{\alpha (\lambda )}\eta ^{(\mu
)}-V_{\alpha }^{\;\;(\mu )}\eta _{(\lambda )}\right) C_{(\mu )}+A_{\alpha
}^{\;\;(\mu )}\eta _{(\lambda )}\eta _{(\mu )}\right)  \nonumber \\
&&+A^{*\alpha (\lambda )}\left( \left( A_{\alpha (\lambda )}C^{(\mu
)}-A_{\alpha }^{\;\;(\mu )}C_{(\lambda )}\right) \eta _{(\mu )}+V_{\alpha
}^{\;\;(\mu )}C_{(\lambda )}C_{(\mu )}\right)  \nonumber \\
&&\left. +\varepsilon ^{\alpha \beta \gamma }\left( A_{\alpha
}^{\;\;(\lambda )}A_{\beta }^{\;\;(\mu )}V_{\gamma (\lambda )}\eta _{(\mu
)}+V_{\alpha }^{\;\;(\lambda )}V_{\beta }^{\;\;(\mu )}A_{\gamma (\lambda
)}C_{(\mu )}\right) \right)  \nonumber \\
&&+\left( 2c_{4}c_{1}^{\prime }-c_{2}c_{4}^{\prime }\right) \left( \left(
V^{*\alpha (\lambda )}\eta _{(\alpha )}+A^{*\lambda (\mu )}C_{(\mu
)}-A_{\;\;\;(\mu )}^{*\mu }C^{(\lambda )}\right) \eta _{(\lambda )}\right.
\nonumber \\
&&\left. +\varepsilon ^{\alpha \lambda \mu }\left( \left( A_{\alpha
}^{\;\;(\beta )}V_{\beta (\lambda )}-A_{\beta }^{\;\;(\beta )}V_{\alpha
(\lambda )}\right) \eta _{(\mu )}+V_{\alpha (\lambda )}V_{\beta (\mu
)}C^{(\beta )}\right) \right)  \nonumber \\
&&+\left( 2c_{3}c_{4}^{\prime }-c_{5}c_{1}^{\prime }\right) \left( \left(
A^{*\alpha (\lambda )}C_{(\alpha )}+V^{*\lambda (\mu )}\eta _{(\mu
)}-V_{\;\;\;(\mu )}^{*\mu }\eta ^{(\lambda )}\right) C_{(\lambda )}\right.
\nonumber \\
&&\left. +\varepsilon ^{\alpha \lambda \mu }\left( \left( V_{\alpha
}^{\;\;(\beta )}A_{\beta (\lambda )}-V_{\beta }^{\;\;(\beta )}A_{\alpha
(\lambda )}\right) C_{(\mu )}+A_{\alpha (\lambda )}A_{\beta (\mu )}\eta
^{(\beta )}\right) \right)  \nonumber \\
&&-\varepsilon ^{\alpha \lambda \mu }\left( c_{2}^{\prime }\left(
c_{1}^{\prime }A_{\alpha (\lambda )}C_{(\mu )}+c_{4}^{\prime }V_{\alpha
(\lambda )}\eta _{(\mu )}\right) \right.  \nonumber \\
&&\left. \left. +\left( \left( c_{2}^{\prime }\right) ^{2}+c_{1}^{\prime
}c_{4}^{\prime }\right) \left( A_{\alpha (\lambda )}\eta _{(\mu )}+V_{\alpha
(\lambda )}C_{(\mu )}\right) \right) \right) .  \label{bf82}
\end{eqnarray}
We observe that none of the terms in the right hand-side of (\ref{bf82}) can
be written like an $s$-exact quantity, such that the consistency of the
first-order deformation requires that these terms must vanish. This takes
place if and only if the constants satisfy the equations
\begin{equation}
4c_{3}c_{4}-c_{2}c_{5}=0,  \label{bf83}
\end{equation}
\begin{equation}
2c_{4}c_{1}^{\prime }-c_{2}c_{4}^{\prime }=0,\;2c_{3}c_{4}^{\prime
}-c_{5}c_{1}^{\prime }=0,  \label{bf84}
\end{equation}
\begin{equation}
c_{2}^{\prime }c_{1}^{\prime }=0,\;c_{2}^{\prime }c_{4}^{\prime }=0,\;\left(
c_{2}^{\prime }\right) ^{2}+c_{1}^{\prime }c_{4}^{\prime }=0.  \label{bf85}
\end{equation}
Consequently, we can take the second-order deformation to vanish, $S_{2}=0$,
and, moreover, we further find that all the higher-order deformations can
also be set equal to zero, $S_{3}=\cdots =0$. In consequence, the
deformation procedure stops at the first order and grants the consistency of
the solution to the master equation at all orders in the coupling constant.
In order to classify the resulting deformations, we solve the Eqs.~(\ref
{bf83}--\ref{bf85}) starting with (\ref{bf85}), that yields
\begin{equation}
c_{2}^{\prime }=0,\;c_{1}^{\prime }c_{4}^{\prime }=0,  \label{bf86}
\end{equation}
so we obtain three distinct possibilities.

\begin{enumerate}
\item[(i)]  The first case corresponds to the vanishing of the solution to
the `homogeneous' equation (\ref{bf68}), $\stackrel{[1]}{a^{\prime }}_{1}=0$%
\begin{equation}
c_{1}^{\prime }=c_{2}^{\prime }=c_{4}^{\prime }=0,  \label{bf87}
\end{equation}
such that (\ref{bf84}) is automatically satisfied. Then, the overall
deformed solution to the master equation, consistent to all orders in the
coupling constant, reads as
\begin{eqnarray}
&&\bar{S}_{(i)}=S+g\varepsilon ^{\lambda \mu \nu }\int d^{3}x\left( \eta
_{(\lambda )}^{*}\left( \frac{c_{5}}{2}\eta _{(\mu )}\eta _{(\nu
)}+c_{2}\eta _{(\mu )}C_{(\nu )}+c_{3}C_{(\mu )}C_{(\nu )}\right) \right.
\nonumber \\
&&+C_{(\lambda )}^{*}\left( \frac{c_{2}}{2}C_{(\mu )}C_{(\nu )}+c_{5}C_{(\mu
)}\eta _{(\nu )}+c_{4}\eta _{(\mu )}\eta _{(\nu )}\right)   \nonumber \\
&&+V_{\;\;\;(\lambda )}^{*\alpha }\left( \left( c_{5}V_{\alpha (\mu
)}+c_{2}A_{\alpha (\mu )}\right) \eta _{(\nu )}+\left( 2c_{3}A_{\alpha (\mu
)}+c_{2}V_{\alpha (\mu )}\right) C_{(\nu )}\right)   \nonumber \\
&&+A_{\;\;\;(\lambda )}^{*\alpha }\left( \left( c_{2}A_{\alpha (\mu
)}+c_{5}V_{\alpha (\mu )}\right) C_{(\nu )}+\left( 2c_{4}V_{\alpha (\mu
)}+c_{5}A_{\alpha (\mu )}\right) \eta _{(\nu )}\right)   \nonumber \\
&&+\varepsilon ^{\alpha \beta \gamma }\left( \frac{1}{2}\left(
c_{2}V_{\alpha (\lambda )}A_{\beta (\mu )}A_{\gamma (\nu )}+c_{5}A_{\alpha
(\lambda )}V_{\beta (\mu )}V_{\gamma (\nu )}\right) \right.   \nonumber \\
&&\left. \left. +\frac{1}{3}\left( c_{3}A_{\alpha (\lambda )}A_{\beta (\mu
)}A_{\gamma (\nu )}+c_{4}V_{\alpha (\lambda )}V_{\beta (\mu )}V_{\gamma (\nu
)}\right) \right) \right) ,  \label{bf88}
\end{eqnarray}
where the four real constants are subject to the equation (\ref{bf83}), and $%
S$ is the free solution (\ref{bf48}).

\item[(ii)]  The second situation is described by
\begin{equation}
c_{1}^{\prime }=c_{2}^{\prime }=0,\;c_{4}^{\prime }\neq 0,  \label{bf89}
\end{equation}
and induces the solution to (\ref{bf83}--\ref{bf84}) like
\begin{equation}
c_{2}=c_{3}=0,  \label{bf90}
\end{equation}
with $c_{4}$ and $c_{5}$ arbitrary real constants. In this case, the full
deformed solution of the master equation is
\begin{eqnarray}
&&\bar{S}_{(ii)}=S+g\int d^{3}x\left( \varepsilon ^{\lambda \mu \nu }\left(
\frac{c_{5}}{2}\eta _{(\lambda )}^{*}\eta _{(\mu )}\eta _{(\nu )}\right.
\right.   \nonumber \\
&&+C_{(\lambda )}^{*}\left( c_{5}C_{(\mu )}\eta _{(\nu )}+c_{4}\eta _{(\mu
)}\eta _{(\nu )}\right) +c_{5}V_{\;\;\;(\lambda )}^{*\alpha }V_{\alpha (\mu
)}\eta _{(\nu )}  \nonumber \\
&&+A_{\;\;\;(\lambda )}^{*\alpha }\left( c_{5}V_{\alpha (\mu )}C_{(\nu
)}+\left( 2c_{4}V_{\alpha (\mu )}+c_{5}A_{\alpha (\mu )}\right) \eta _{(\nu
)}\right)   \nonumber \\
&&\left. +\varepsilon ^{\alpha \beta \gamma }\left( \frac{c_{5}}{2}A_{\alpha
(\lambda )}V_{\beta (\mu )}V_{\gamma (\nu )}+\frac{c_{4}}{3}V_{\alpha
(\lambda )}V_{\beta (\mu )}V_{\gamma (\nu )}\right) \right)   \nonumber \\
&&\left. +c_{4}^{\prime }\left( \varepsilon ^{\alpha \lambda \mu }A_{\alpha
(\lambda )}^{*}\eta _{(\mu )}+\frac{1}{2}\left( V_{\;\;(\beta )}^{\gamma
}V_{\;\;(\gamma )}^{\beta }-V_{\;\;(\beta )}^{\beta }V_{\;\;(\gamma
)}^{\gamma }\right) \right) \right) .  \label{bf91}
\end{eqnarray}

\item[(iii)]  Finally, in the third case we have that
\begin{equation}
c_{2}^{\prime }=c_{4}^{\prime }=0,\;c_{1}^{\prime }\neq 0,  \label{bf92}
\end{equation}
which further leads to the solution of (\ref{bf83}--\ref{bf84}) under the
form
\begin{equation}
c_{4}=c_{5}=0,  \label{bf93}
\end{equation}
with $c_{2}$ and $c_{3}$ arbitrary real constants. In this situation, the
consistent deformed solution of the master equation is expressed by
\begin{eqnarray}
&&\bar{S}_{(iii)}=S+g\int d^{3}x\left( \varepsilon ^{\lambda \mu \nu }\left(
\frac{c_{2}}{2}C_{(\lambda )}^{*}C_{(\mu )}C_{(\nu )}\right. \right.
\nonumber \\
&&+\eta _{(\lambda )}^{*}\left( c_{2}\eta _{(\mu )}C_{(\nu )}+c_{3}C_{(\mu
)}C_{(\nu )}\right) +c_{2}A_{\;\;\;(\lambda )}^{*\alpha }A_{\alpha (\mu
)}C_{(\nu )}  \nonumber \\
&&+V_{\;\;\;(\lambda )}^{*\alpha }\left( c_{2}A_{\alpha (\mu )}\eta _{(\nu
)}+\left( 2c_{3}A_{\alpha (\mu )}+c_{2}V_{\alpha (\mu )}\right) C_{(\nu
)}\right)   \nonumber \\
&&\left. +\varepsilon ^{\alpha \beta \gamma }\left( \frac{c_{2}}{2}V_{\alpha
(\lambda )}A_{\beta (\mu )}A_{\gamma (\nu )}+\frac{c_{3}}{3}A_{\alpha
(\lambda )}A_{\beta (\mu )}A_{\gamma (\nu )}\right) \right)   \nonumber \\
&&\left. +c_{1}^{\prime }\left( \varepsilon ^{\alpha \lambda \mu }V_{\alpha
(\lambda )}^{*}C_{(\mu )}+\frac{1}{2}\left( A_{\;\;(\beta )}^{\gamma
}A_{\;\;(\gamma )}^{\beta }-A_{\;\;(\beta )}^{\beta }A_{\;\;(\gamma
)}^{\gamma }\right) \right) \right) .  \label{bf94}
\end{eqnarray}
\end{enumerate}

It is interesting to notice that in each situation the deformed solution to
the master equation is parametrized in terms of three independent real
constants.

\section{Identification of the Interacting Model}

We are now in the position to identify the resulting interaction models. As
we have previously mentioned, the antighost number zero component in the
deformed solution to the master equation determines the Lagrangian action of
the coupled model. From the antighost number one elements we read the
deformed gauge transformations. The pieces with higher antighost number
offer information on the nature of the deformed gauge algebra. In the first
of the cases discussed in the above, we infer the deformed Lagrangian action
and accompanying gauge transformations under the form
\begin{eqnarray}
&&I_{(i)}\left[ V_{\alpha (\lambda )},A_{\alpha }^{\;\;(\lambda )}\right]
=\varepsilon ^{\alpha \beta \gamma }\int d^{3}x\left( V_{\alpha (\lambda
)}\partial _{\beta }A_{\gamma }^{\;\;(\lambda )}\right.  \nonumber \\
&&+g\varepsilon ^{\lambda \mu \nu }\left( \frac{1}{2}\left( c_{2}V_{\alpha
(\lambda )}A_{\beta (\mu )}A_{\gamma (\nu )}+c_{5}A_{\alpha (\lambda
)}V_{\beta (\mu )}V_{\gamma (\nu )}\right) \right.  \nonumber \\
&&\left. \left. +\frac{1}{3}\left( c_{3}A_{\alpha (\lambda )}A_{\beta (\mu
)}A_{\gamma (\nu )}+c_{4}V_{\alpha (\lambda )}V_{\beta (\mu )}V_{\gamma (\nu
)}\right) \right) \right) ,  \label{bf95}
\end{eqnarray}
\begin{eqnarray}
&&\bar{\delta}_{\epsilon }^{(i)}V^{\alpha (\lambda )}=\partial ^{\alpha
}\epsilon ^{\prime (\lambda )}+g\varepsilon ^{\lambda \mu \nu }\left( \left(
c_{5}V_{\;\;(\mu )}^{\alpha }+c_{2}A_{\;\;(\mu )}^{\alpha }\right) \epsilon
_{(\nu )}^{\prime }\right.  \nonumber \\
&&\left. +\left( 2c_{3}A_{\;\;(\mu )}^{\alpha }+c_{2}V_{\;\;(\mu )}^{\alpha
}\right) \epsilon _{(\nu )}\right) ,  \label{bf96}
\end{eqnarray}
\begin{eqnarray}
&&\bar{\delta}_{\epsilon }^{(i)}A^{\alpha (\lambda )}=\partial ^{\alpha
}\epsilon ^{(\lambda )}+g\varepsilon ^{\lambda \mu \nu }\left( \left(
c_{2}A_{\;\;(\mu )}^{\alpha }+c_{5}V_{\;\;(\mu )}^{\alpha }\right) \epsilon
_{(\nu )}\right.  \nonumber \\
&&\left. +\left( 2c_{4}V_{\;\;(\mu )}^{\alpha }+c_{5}A_{\;\;(\mu )}^{\alpha
}\right) \epsilon _{(\nu )}^{\prime }\right) .  \label{bf97}
\end{eqnarray}
We remark that the above deformed action, as well as its gauge
transformations, preserve the symmetry under the transformations (\ref{bf81}%
--\ref{bf80}).

In the second and third cases, we infer that
\begin{eqnarray}
&&I_{(ii)}\left[ V_{\alpha (\lambda )},A_{\alpha }^{\;\;(\lambda )}\right]
=\int d^{3}x\left( \varepsilon ^{\alpha \beta \gamma }\left( V_{\alpha
(\lambda )}\partial _{\beta }A_{\gamma }^{\;\;(\lambda )}\right. \right.
\nonumber \\
&&\left. +g\varepsilon ^{\lambda \mu \nu }\left( \frac{c_{5}}{2}A_{\alpha
(\lambda )}V_{\beta (\mu )}V_{\gamma (\nu )}+\frac{c_{4}}{3}V_{\alpha
(\lambda )}V_{\beta (\mu )}V_{\gamma (\nu )}\right) \right)  \nonumber \\
&&\left. +g\frac{c_{4}^{\prime }}{2}\left( V_{\;\;(\beta )}^{\gamma
}V_{\;\;(\gamma )}^{\beta }-V_{\;\;(\beta )}^{\beta }V_{\;\;(\gamma
)}^{\gamma }\right) \right) ,  \label{bf98}
\end{eqnarray}
\begin{equation}
\bar{\delta}_{\epsilon }^{(ii)}V^{\alpha (\lambda )}=\partial ^{\alpha
}\epsilon ^{\prime (\lambda )}+gc_{5}\varepsilon ^{\lambda \mu \nu
}V_{\;\;(\mu )}^{\alpha }\epsilon _{(\nu )}^{\prime },  \label{bf99}
\end{equation}
\begin{eqnarray}
&&\bar{\delta}_{\epsilon }^{(ii)}A^{\alpha (\lambda )}=\partial ^{\alpha
}\epsilon ^{(\lambda )}+gc_{4}^{\prime }\varepsilon ^{\alpha \lambda \mu
}\epsilon _{(\mu )}^{\prime }  \nonumber \\
&&+g\varepsilon ^{\lambda \mu \nu }\left( c_{5}V_{\;\;(\mu )}^{\alpha
}\epsilon _{(\nu )}+\left( 2c_{4}V_{\;\;(\mu )}^{\alpha }+c_{5}A_{\;\;(\mu
)}^{\alpha }\right) \epsilon _{(\nu )}^{\prime }\right) ,  \label{bf100}
\end{eqnarray}
respectively,
\begin{eqnarray}
&&I_{(iii)}\left[ V_{\alpha (\lambda )},A_{\alpha }^{\;\;(\lambda )}\right]
=\int d^{3}x\left( \varepsilon ^{\alpha \beta \gamma }\left( V_{\alpha
(\lambda )}\partial _{\beta }A_{\gamma }^{\;\;(\lambda )}\right. \right.
\nonumber \\
&&\left. +g\varepsilon ^{\lambda \mu \nu }\left( \frac{c_{2}}{2}V_{\alpha
(\lambda )}A_{\beta (\mu )}A_{\gamma (\nu )}+\frac{c_{3}}{3}A_{\alpha
(\lambda )}A_{\beta (\mu )}A_{\gamma (\nu )}\right) \right)  \nonumber \\
&&\left. +g\frac{c_{1}^{\prime }}{2}\left( A_{\;\;(\beta )}^{\gamma
}A_{\;\;(\gamma )}^{\beta }-A_{\;\;(\beta )}^{\beta }A_{\;\;(\gamma
)}^{\gamma }\right) \right) ,  \label{bf101}
\end{eqnarray}
\begin{eqnarray}
&&\bar{\delta}_{\epsilon }^{(iii)}V^{\alpha (\lambda )}=\partial ^{\alpha
}\epsilon ^{\prime (\lambda )}+gc_{1}^{\prime }\varepsilon ^{\alpha \lambda
\mu }\epsilon _{(\mu )}  \nonumber \\
&&+g\varepsilon ^{\lambda \mu \nu }\left( c_{2}A_{\;\;(\mu )}^{\alpha
}\epsilon _{(\nu )}^{\prime }+\left( 2c_{3}A_{\;\;(\mu )}^{\alpha
}+c_{2}V_{\;\;(\mu )}^{\alpha }\right) \epsilon _{(\nu )}\right) ,
\label{bf102}
\end{eqnarray}
\begin{equation}
\bar{\delta}_{\epsilon }^{(iii)}A^{\alpha (\lambda )}=\partial ^{\alpha
}\epsilon ^{(\lambda )}+gc_{2}\varepsilon ^{\lambda \mu \nu }A_{\;\;(\mu
)}^{\alpha }\epsilon _{(\nu )}.  \label{bf103}
\end{equation}
Related to the last two situations, we notice that they can be obtained one
from the other by performing the transformations (\ref{bf81}--\ref{bf80}).
In this sense, the cases (ii) and (iii) are complementary.

By inspecting the structure of the terms with antighost number two appearing
in each of the solutions (\ref{bf88}), (\ref{bf91}) and (\ref{bf94}), we
observe that in all cases the gauge algebra is deformed and closes according
to a $su(2)$ Lie algebra, with the structure constants proportional with the
components of the completely antisymmetric symbol in three dimensions, $%
\varepsilon ^{\lambda \mu \nu }$.

Finally, we investigate the particular case where the degree three tensor
gauge field $B_{\;\;\;(\lambda )}^{\alpha \beta }$ is dual to that of degree
two with respect to the antisymmetry indices only, $A_{\alpha (\lambda
)}=\varepsilon _{\alpha \beta \gamma }B_{\;\;\;(\lambda )}^{\beta \gamma }$,
which is translated into $A_{\alpha (\lambda )}=V_{\alpha (\lambda )}$ at
the level of action (\ref{bf33}) and of its gauge symmetries (\ref{bf34}).
In this situation, we start from the Lagrangian action and associated gauge
transformations
\begin{equation}
I_{0}\left[ A_{\alpha }^{\;\;(\lambda )}\right] =\int d^{3}x\varepsilon
^{\alpha \beta \gamma }A_{\alpha (\lambda )}\partial _{\beta }A_{\gamma
}^{\;\;(\lambda )},\;\delta _{\epsilon }A_{\alpha }^{\;\;(\lambda
)}=\partial _{\alpha }\epsilon ^{(\lambda )},  \label{bf104}
\end{equation}
such that the free solution to the master equation (\ref{bf48}) becomes
\begin{equation}
S=I_{0}\left[ A_{\alpha }^{\;\;(\lambda )}\right] +\int
d^{3}x\,A_{\;\;\;(\lambda )}^{*\alpha }\partial _{\alpha }C^{(\lambda )}.
\label{bf105}
\end{equation}
The analogue of the first-order deformation (\ref{bf79}) is found of the
type
\begin{eqnarray}
&&S_{1}=\int d^{3}x\left( c\varepsilon ^{\lambda \mu \nu }\left( \frac{1}{2}%
C_{(\lambda )}^{*}C_{(\mu )}C_{(\nu )}+A_{\;\;\;(\lambda )}^{*\alpha
}A_{\alpha (\mu )}C_{(\nu )}\right. \right.  \nonumber \\
&&\left. +\frac{1}{3}\varepsilon ^{\alpha \beta \gamma }A_{\alpha (\lambda
)}A_{\beta (\mu )}A_{\gamma (\nu )}\right)  \nonumber \\
&&\left. +c^{\prime }\left( \varepsilon ^{\alpha \lambda \mu }A_{\alpha
(\lambda )}^{*}C_{(\mu )}+\left( A_{\;\;(\beta )}^{\gamma }A_{\;\;(\gamma
)}^{\beta }-A_{\;\;(\beta )}^{\beta }A_{\;\;(\gamma )}^{\gamma }\right)
\right) \right) ,  \label{bf106}
\end{eqnarray}
while its consistency, (\ref{bf54}), holds if and only if
\begin{equation}
c^{\prime }=0,  \label{bf107}
\end{equation}
while $c$ remains arbitrary, and can be fixed equal to the unity, $c=1$. The
deformed Lagrangian action and gauge symmetries will accordingly be
\begin{equation}
I\left[ A_{\alpha }^{\;\;(\lambda )}\right] =\varepsilon ^{\alpha \beta
\gamma }\int d^{3}xA_{\alpha (\lambda )}\left( \partial _{\beta }A_{\gamma
}^{\;\;(\lambda )}+\frac{g}{3}\varepsilon ^{\lambda \mu \nu }A_{\beta (\mu
)}A_{\gamma (\nu )}\right) ,  \label{bf108}
\end{equation}
\begin{equation}
\bar{\delta}_{\epsilon }A_{\alpha }^{\;\;(\lambda )}=\partial _{\alpha
}\epsilon ^{(\lambda )}+g\varepsilon ^{\lambda \mu \nu }A_{\alpha (\mu
)}\epsilon _{(\nu )},  \label{bf109}
\end{equation}
such that the deformed gauge algebra is again $su(2)$. This particular case
is similar to what happens during the deformation of pure three-dimensional
Abelian Chern--Simons theory into the non-Abelian version.

\section{Conclusion}

In conclusion, in this work we have constructed all consistent Lagrangian
interactions that can be added to a free theory involving tensor gauge
fields of degrees two and three in three space-time dimensions with the help
of the free local BRST cohomology. We have proven that the deformed solution
to the master equation can be taken to be non-vanishing only at the first
order in the coupling constant. As a consequence, we determine three types
of interacting models (among which two are complementary in the sense that
they can be obtained one from the other by performing a certain
transformation) with deformed gauge transformations, a non-Abelian ($su(2)$)
gauge algebra and polynomial vertices of order three in the undifferentiated
fields. In the special case where the tensor gauge fields are dual to each
other, we obtain an analogue of pure three-dimensional Chern--Simons theory.

\section*{Acknowledgments}

This work has been supported by a type-A grant with CNCSIS-MEC, Romania.

\end{document}